# ALICE ITS3: how to integrate a large dimension MAPS sensor in a bent configuration detector

**Domenico Colella** [a],[*] **for the ALICE Collaboration**

[a] *University and INFN Bari,*
 *Via E. Orabona 4, 170125, Bari, Italy*

 *E-mail:* domenico.colella@ba.infn.it

The ALICE Collaboration is developing a novel vertexing detector to extend the heavy-flavour physics programme of the experiment during Run 4 by improving the pointing resolution of the tracking, particularly at low transverse momentum. It will be a detector with three truly cylindrical layers based on thin wafer scale MAPS, reaching less than 0.07% $X/X_0$ per layer and with the innermost layer located as close as 19 mm to the interaction point. This contribution will describe the global detector integration concept, focusing on: the sensor bending procedure, the sensor electrical interconnection, the choice of the best carbon foam for light mechanical supporting structures, the studies of cooling by air flow, and the global structures mechanical characterization.

## 1. Introduction

The ITS3 project of the ALICE Collaboration [1,2], is developing an ultra-light vertexing detector, combining several innovative approaches, many of which have never been adopted in high-energy physics. The result is a nearly massless detector that is based on wafer-scale silicon sensors of up to 10 cm × 26.6 cm, reaching a mean material thickness of 0.09% $X/X_0$ per layer.

Two new pioneering technologies are implemented in the sensor design: usage of 65 nm CMOS technology node that allows more dense circuitry and gives access to larger wafers (300 mm in diameter), and stitching technique that allows for extending the sensor dimentions above the usual size ($O(2\times3\ cm^2)$). The implementation of the main analog and digital blocks of a MAPS in 65 nm has been demostrated with a large campaign of sensor characterization that took place from 2020 to 2022; results are given in Ref. [3]. The development of large stitched sensors took two years of design and led to reception of first prototypes during summer 2023; the results of first characterization are given in Ref. [4].

The other groundbreaking innovation is the removal of most of the services from the detector acceptance exploting the flexibility of thin silicon. Detector half-layers will be formed by a single sensor, thinned down to 50 µm, bent to acquire a half-cylindrical shape, and fixed in position by carbon foam structures. Sensor cooling will be performed by forced air flow between detector layers and by thermal conduction to the support structures.

## 2. Detector concept

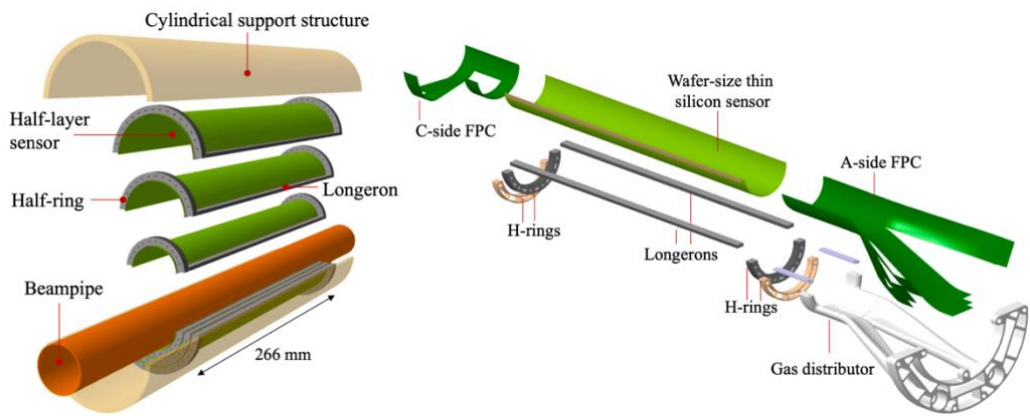

**Figure 1**: Left: ITS3 detector concept with six half-layers. Right: components of single half-layer. Taken from Ref. [2].

The ITS3 detector is structured into three layers, with respective target radii of 19.0 mm, 25.2 mm, and 31.5 mm. Each layer is further divided into two halves, each forming a half-cylinder. The three half-layers are in contact through the support structures and are mounted in an external cylindrical support structure, forming a compact block to be placed around the beam pipe. A simplified schematic is shown in the left part of Figure 1. A more detailed half-layer schematic, including services like air distributor and flex for powering, control, and data transmission, is shown in the right part of Figure 1. The whole detector will thus be made of only six sensors, each of them covering 180 degrees in azimuthal angle, and measuring 266 mm in length. These six sensors will be kept in cylindrical shape by minimal ultra-light support structures made of carbon foam, placed at the edges of the sensor sensible area. Air for cooling will be driven to the sensor through small diameter ducts. These ducts will also be a part of the support structure for the flex circuitry that will be used to bring power and allow comunication with sensor. As can be seen in Figure 1, all these services will lie outside the detector sensible area.

### 3. Sensor bending

R&D activities started from small dimension sensors. A 3-point bending test, demonstrated that ITS 3 target radii are easily reachable with silicon thickness below 50 μm, for sensors based on CMOS technology node of 180 nm and 65 nm [2]. It has also been shown that MAPS (180 nm node) do not change their performance when bent to acquire cylindrical shape withthe ITS 3 target radii [5].

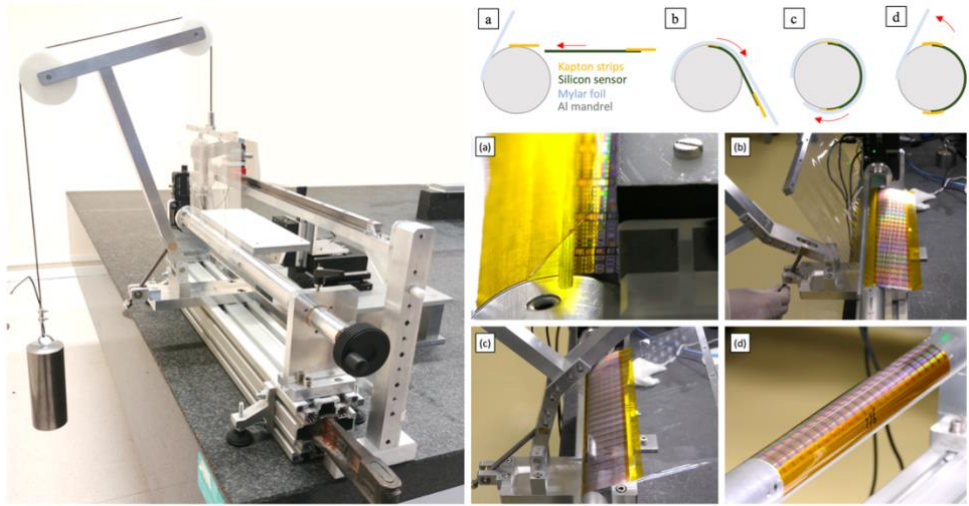

**Figure 2**: Left: picture of the bending setup for a large dimension silicon sensor. Right: details of the bending procedure steps. Taken from Ref. [2].

Large dimension silicon sensors are bent in cylindrical shape by means of a mandrel and a tensioned mylar foil. After the alignment to the mandrel, an adhesive polyimide strip holds the sensor to the mandrel. A rotating motor wraps the mylar foil around the mandrel forcing the sensor to bend. A second adhesive polyimide strip guarantees that the sensor will be kept in position at the end of the procedure. Figure 2 shows the bending tool and the main procedure steps. Such a bending procedure is nowadays routinely performed with success.

### 4. Mechanics and cooling

Two main components are used as support structures for the half-layer: the logerons and the half-rings (Figure 1). Carbon foam has been identified as the optimal material to satisfy various requirements for the ITS3. Two varieties of open-cell reticulated vitreous carbon (RVC) foams have been chosen based on the functionality of the mechanical components. A low-density foam (45 kg/m$^3$) with a high stiffness, Carbon Duocel, is selected for the support longerons. Meanwhile, for the cooling radiator, the Allcomp K9 standard density RVC foam, distinguished by its high thermal conductivity (25 W m$^{-1}$ K$^{-1}$), has been chosen.

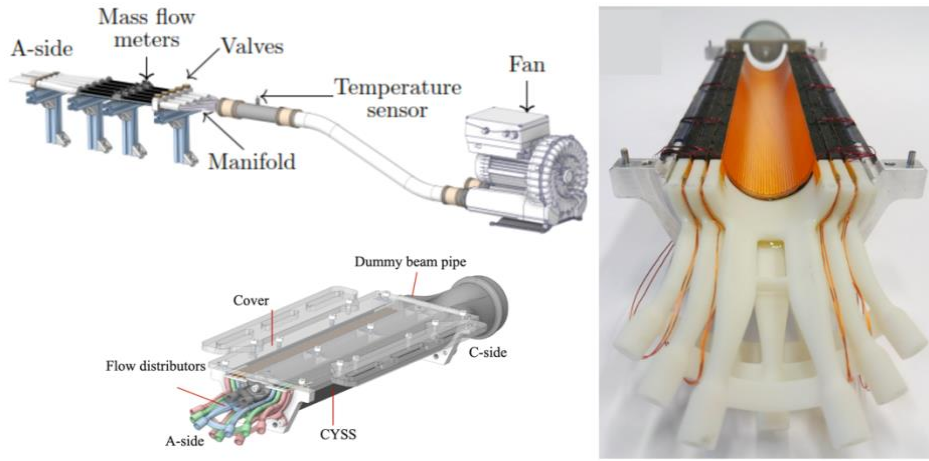

**Figure 3**: Left: wind tunnel setup for air cooling verification. Right: ITS3 prototype integrating heaters to simulate power dissipation. Taken from Ref. [2].

A prototype, integrating ducts for air cooling and heaters to simulate the power dissipation of the MAPS, has been used in a wind-tunnel setup (Figure 3), to verify the possibility of sensor cooling by air flow. Power dissipation of the sensor will not be uniform across the silicon surface, so the prototype integrated two different regions: the matrix (25 mW/cm$^2$) and end-cap regions (1000 mW/cm$^2$). It has been demonstrated that, with an average freestream airflow velocity of approximately 8 m/s between the layers, the detector is kept at a temperature that is 5 °C higher than the inlet air temperature and uniformity is maintained within 5 °C along the sensor.

Further mechanical characterizations of the half-detector established that the maximum displacement of a sensor, due to the airflow is limited to about 1.1 µm with an RMS below 0.4 µm, which well meets the ITS 3 requirements.

Finally, another prototype, made of a single silicon layer glued to a carbon fiber cylindrical external structure, has been tested in a climate chamber (with temperature ranging between 10 and 40 °C and constant 50% humidity), showing no detrimental effects from the temperature fluctuations. This is the first measurement of a larger campaign that will assess potential failures caused by thermo-elastic deformation.

### 5. Flex printed circuit integration

Sensor powering will be provided at the two longitudinal extremities by two flex printed circuits (FPC). A-side FPC (Figure 1) is also designed to perform the communication (control and data transmission) to the external world. This specific FPC is designed to have a region bent at the same radius of the corresponding sensor, and tails having zero insertion force connectors to the external boards. Length of the full FPC is limited by signal transmission integrity to less than 30 cm. Electrical connection to the sensor is done via wire-bonding. A safety gap of 0.5 mm is left between the sensor and the FPC edges, reflecting in a distance between corresponding pads for soldering in the two objects of the order of 1 mm. A first exercise to estimate the best bonding parameters, targeting the best pull-force and minimizing the loop height, returned a mean pull-force value of 6.6±0.3 g. This study, done with not final grade materials, will be repeated with final components.

### 6. Summary

The intense R&D activities done in the ITS 3 project already successfully demonstrated the functionality of first stitched sensor and defined a controlled procedure to bend wafer-scale silicon sensors. Next sensor submission is planned for 2024, containing the prototype sensor to be used in the ALICE ITS 3 detector. On mechanical side more studies are expected to further improve the cooling system and assess the mechanical properties of the half-barrel, providing guidelines for the final detector assembly procedure. The final detector is expected to be assembled during the first quarter of 2027, to be installed and finally commissioned in the cavern during the first quarter of 2028.